\documentclass[aps,prd,reprint,showpacs,showkeys,nofootinbib,superscriptaddress]{revtex4-1}

\usepackage{graphicx}
\usepackage{epstopdf}

\newcommand{\kgd}{kg$\cdot$d~}
\newcommand{\kgdd}{kg$\cdot$d}

\begin{document}

\title{A search for low-mass WIMPs
with EDELWEISS-II heat-and-ionization detectors}

\author{E.~Armengaud}
\email{eric.armengaud@cea.fr}
\affiliation{CEA, Centre d'Etudes Saclay, IRFU, 91191 Gif-Sur-Yvette Cedex, France}
\author{C.~Augier}
\affiliation{IPNL, Universit\'{e} de Lyon, Universit\'{e} Lyon 1, CNRS/IN2P3, 4 rue E. Fermi 69622 Villeurbanne cedex, France}
\author{A.~Beno\^{\i}t}
\affiliation{CNRS-N\'{e}el, 25 Avenue des Martyrs, 38042 Grenoble cedex 9, France}
\author{L.~Berg\'e}
\affiliation{CSNSM, Universit\'e Paris-Sud, IN2P3-CNRS, bat 108, 91405 Orsay,  France}
\author{T.~Bergmann}
\affiliation{Karlsruhe Institute of Technology, Institut f\"ur Prozessdatenverarbeitung und Elektronik, 76021 Karlsruhe, Germany}
\author{J.~Bl$\mbox{\"u}$mer}
\affiliation{Karlsruhe Institute of Technology, Institut f\"ur Experimentelle Kernphysik, 76128 Karlsruhe, Germany}
\affiliation{Karlsruhe Institute of Technology, Institut f\"ur Kernphysik, 76021 Karlsruhe, Germany}
\author{A.~Broniatowski}
\affiliation{CSNSM, Universit\'e Paris-Sud, IN2P3-CNRS, bat 108, 91405 Orsay,  France}
\author{V.~Brudanin}
\affiliation{Laboratory of Nuclear Problems, JINR, Joliot-Curie 6, 141980 Dubna, Moscow region, Russia}
\author{B.~Censier}
\affiliation{IPNL, Universit\'{e} de Lyon, Universit\'{e} Lyon 1, CNRS/IN2P3, 4 rue E. Fermi 69622 Villeurbanne cedex, France}
\author{M.~Chapellier}
\affiliation{CSNSM, Universit\'e Paris-Sud, IN2P3-CNRS, bat 108, 91405 Orsay,  France}
\author{F.~Charlieux}
\affiliation{IPNL, Universit\'{e} de Lyon, Universit\'{e} Lyon 1, CNRS/IN2P3, 4 rue E. Fermi 69622 Villeurbanne cedex, France}
\author{F.~Cou\"{e}do}
\affiliation{CSNSM, Universit\'e Paris-Sud, IN2P3-CNRS, bat 108, 91405 Orsay,  France}
\author{P.~Coulter}
\affiliation{University of Oxford, Department of Physics, Keble Road, Oxford OX1 3RH, UK}
\author{G.A.~Cox}
\affiliation{Karlsruhe Institute of Technology, Institut f\"ur Experimentelle Kernphysik, 76128 Karlsruhe, Germany}
\author{J.~Domange}
\affiliation{CSNSM, Universit\'e Paris-Sud, IN2P3-CNRS, bat 108, 91405 Orsay,  France}
\affiliation{CEA, Centre d'Etudes Saclay, IRFU, 91191 Gif-Sur-Yvette Cedex, France}
\author{A.A.~Drillien}
\affiliation{CSNSM, Universit\'e Paris-Sud, IN2P3-CNRS, bat 108, 91405 Orsay,  France}
\author{L.~Dumoulin}
\affiliation{CSNSM, Universit\'e Paris-Sud, IN2P3-CNRS, bat 108, 91405 Orsay,  France}
\author{K.~Eitel}
\affiliation{Karlsruhe Institute of Technology, Institut f\"ur Kernphysik, 76021 Karlsruhe, Germany}
\author{D.~Filosofov}
\affiliation{Laboratory of Nuclear Problems, JINR, Joliot-Curie 6, 141980 Dubna, Moscow region, Russia}
\author{N.~Fourches}
\affiliation{CEA, Centre d'Etudes Saclay, IRFU, 91191 Gif-Sur-Yvette Cedex, France}
\author{J.~Gascon}
\affiliation{IPNL, Universit\'{e} de Lyon, Universit\'{e} Lyon 1, CNRS/IN2P3, 4 rue E. Fermi 69622 Villeurbanne cedex, France}
\author{G.~Gerbier}
\affiliation{CEA, Centre d'Etudes Saclay, IRFU, 91191 Gif-Sur-Yvette Cedex, France}
\author{J.~Gironnet}
\affiliation{IPNL, Universit\'{e} de Lyon, Universit\'{e} Lyon 1, CNRS/IN2P3, 4 rue E. Fermi 69622 Villeurbanne cedex, France}
\author{M.~Gros}
\affiliation{CEA, Centre d'Etudes Saclay, IRFU, 91191 Gif-Sur-Yvette Cedex, France}
\author{S.~Henry}
\affiliation{University of Oxford, Department of Physics, Keble Road, Oxford OX1 3RH, UK}
\author{G.~Heuermann}
\affiliation{Karlsruhe Institute of Technology, Institut f\"ur Experimentelle Kernphysik, 76128 Karlsruhe, Germany}
\author{S.~Herv\mbox{\'e}}
\affiliation{CEA, Centre d'Etudes Saclay, IRFU, 91191 Gif-Sur-Yvette Cedex, France}
\author{A.~Juillard}
\affiliation{IPNL, Universit\'{e} de Lyon, Universit\'{e} Lyon 1, CNRS/IN2P3, 4 rue E. Fermi 69622 Villeurbanne cedex, France}
\author{M.~Kleifges}
\affiliation{Karlsruhe Institute of Technology, Institut f\"ur Prozessdatenverarbeitung und Elektronik, 76021 Karlsruhe, Germany}
\author{H.~Kluck}
\affiliation{Karlsruhe Institute of Technology, Institut f\"ur Experimentelle Kernphysik, 76128 Karlsruhe, Germany}
\author{V.~Kozlov}
\affiliation{Karlsruhe Institute of Technology, Institut f\"ur Kernphysik, 76021 Karlsruhe, Germany}
\author{H.~Kraus}
\affiliation{University of Oxford, Department of Physics, Keble Road, Oxford OX1 3RH, UK}
\author{V.A.~Kudryavtsev}
\affiliation{Department of Physics and Astronomy, University of Sheffield, Hounsfield Road, Sheffield S3 7RH, UK}
\author{H.~Le Sueur}
\affiliation{CSNSM, Universit\'e Paris-Sud, IN2P3-CNRS, bat 108, 91405 Orsay,  France}
\author{P.~Loaiza}
\affiliation{Laboratoire Souterrain de Modane, CEA-CNRS, 1125 route de Bardonn\`eche, 73500 Modane, France}
\author{S.~Marnieros}
\affiliation{CSNSM, Universit\'e Paris-Sud, IN2P3-CNRS, bat 108, 91405 Orsay,  France}
\author{A.~Menshikov}
\affiliation{Karlsruhe Institute of Technology, Institut f\"ur Prozessdatenverarbeitung und Elektronik, 76021 Karlsruhe, Germany}
\author{X-F.~Navick}
\affiliation{CEA, Centre d'Etudes Saclay, IRFU, 91191 Gif-Sur-Yvette Cedex, France}
\author{C.~Nones}
\affiliation{CEA, Centre d'Etudes Saclay, IRFU, 91191 Gif-Sur-Yvette Cedex, France}
\author{E.~Olivieri}
\affiliation{CSNSM, Universit\'e Paris-Sud, IN2P3-CNRS, bat 108, 91405 Orsay,  France}
\author{P.~Pari}
\affiliation{CEA, Centre d'Etudes Saclay, IRAMIS, 91191 Gif-Sur-Yvette Cedex, France}
\author{B.~Paul}
\affiliation{CEA, Centre d'Etudes Saclay, IRFU, 91191 Gif-Sur-Yvette Cedex, France}
\author{M. Robinson}
\affiliation{Department of Physics and Astronomy, University of Sheffield, Hounsfield Road, Sheffield S3 7RH, UK}
\author{S.~Rozov}
\affiliation{Laboratory of Nuclear Problems, JINR, Joliot-Curie 6, 141980 Dubna, Moscow region, Russia}
\author{V.~Sanglard}
\affiliation{IPNL, Universit\'{e} de Lyon, Universit\'{e} Lyon 1, CNRS/IN2P3, 4 rue E. Fermi 69622 Villeurbanne cedex, France}
\author{B.~Schmidt}
\affiliation{Karlsruhe Institute of Technology, Institut f\"ur Experimentelle Kernphysik, 76128 Karlsruhe, Germany}
\author{B.~Siebenborn}
\affiliation{Karlsruhe Institute of Technology, Institut f\"ur Experimentelle Kernphysik, 76128 Karlsruhe, Germany}
\author{D.~Tcherniakhovski}
\affiliation{Karlsruhe Institute of Technology, Institut f\"ur Prozessdatenverarbeitung und Elektronik, 76021 Karlsruhe, Germany}
\author{A.S.~Torrento-Coello}
\affiliation{CEA, Centre d'Etudes Saclay, IRFU, 91191 Gif-Sur-Yvette Cedex, France}
\author{L.~Vagneron}
\affiliation{IPNL, Universit\'{e} de Lyon, Universit\'{e} Lyon 1, CNRS/IN2P3, 4 rue E. Fermi 69622 Villeurbanne cedex, France}
\author{R.J.~Walker}
\affiliation{CEA, Centre d'Etudes Saclay, IRFU, 91191 Gif-Sur-Yvette Cedex, France}
\author{M.~Weber}
\affiliation{Karlsruhe Institute of Technology, Institut f\"ur Prozessdatenverarbeitung und Elektronik, 76021 Karlsruhe, Germany}
\author{E.~Yakushev}
\affiliation{Laboratory of Nuclear Problems, JINR, Joliot-Curie 6, 141980 Dubna, Moscow region, Russia}
\author{X.~Zhang}
\affiliation{University of Oxford, Department of Physics, Keble Road, Oxford OX1 3RH, UK}

\collaboration{EDELWEISS Collaboration}
\noaffiliation

\begin{abstract}
We report on a search for low-energy ($E < 20$~keV) WIMP-induced nuclear recoils using data collected in $2009-2010$ by EDELWEISS from four germanium detectors equipped with thermal sensors and an electrode design (ID) which allows to efficiently reject several sources of background. The data indicate no evidence for an exponential distribution of low-energy nuclear recoils that could be attributed to WIMP elastic scattering after an exposure of 113~\kgdd. For WIMPs of mass 10~GeV, the observation of one event in the WIMP search region results in a 90\%~CL limit of $1.0\times10^{-5}$~pb on the spin-independent WIMP-nucleon scattering cross-section, which constrains the parameter space associated with the findings reported by the CoGeNT, DAMA and CRESST experiments.
\end{abstract}


\maketitle

\section{Introduction}

Within the current cosmological concordance model, a large fraction of the mass content of the observable universe is made of dark matter, as demonstrated by a range of observations but the nature of which is still unknown~\cite{dmreview}. In particular, the dynamics of our galaxy can be explained by the presence of a dark matter halo with a solar neighborhood density of $0.3\pm0.1$~GeV/cm$^3$. Weakly Interacting Massive Particles (WIMPs) constitute a generic class of dark matter candidates. The calculated thermal relic density of WIMPs roughly matches the measured density on cosmological scales. Although several constrained WIMP models point to WIMP masses $M_{\chi}\sim 100$~GeV, lighter fermions with masses down to $\sim 2$~GeV are possible~\cite{leeweinberg,bottino}.

The direct detection of low-mass ($\sim$10 GeV or lower) WIMPs is challenging because the recoil energies generated by the elastic scattering on nuclei of such low-mass particles are close to the experimental thresholds of existing detectors. Experiments using a range of technologies have either constrained~\cite{cdms-lowmass, xe-lowmass} or found hints~\cite{dama,cogent,cresst} of such WIMPs with cross sections for spin-independent scattering on a nucleon $\sigma_{\rm SI}$ of the order of magnitude of $10^{-4}$~pb. There is a discussion regarding these results due to uncertainties on efficiencies, energy scales and residual backgrounds at low recoil energies (see for example~\cite{collar}).

In this article we present a search for low-mass WIMPs using data collected by the EDELWEISS-II detectors in $2009-2010$. 
A WIMP search optimized for WIMP masses above 50~GeV has already been published~\cite{idfirst,edw2,edwcdms}, where data from ten heat-and-ionization germanium detectors with interleaved electrodes were used.
The analysis threshold was set at a recoil energy of 20~keV in order to ensure a maximum exposure in a recoil energy range where the behavior of all of the ten detectors was homogeneous and well understood, both in terms of expected background and rejection capabilities. 
This strategy, however, is not adequate for a search dedicated to models in which the WIMP mass is of the order of 10 GeV, for which the highest expected recoil energy is of the order of 10~keV. In the study presented here, we use a restricted data set, selected on the basis of detector thresholds and backgrounds, for which a low-background sensitivity to nuclear recoils down to 5~keV could be achieved. By limiting this analysis to energies less than 20~keV, the results presented in this article are completely independent from the previously published limits~\cite{edw2}.
After a presentation of data selection, we discuss the potential residual backgrounds in the WIMP search region.  A total exposure of 113~\kgd is obtained, from which we derive constraints on $\sigma_{\rm SI}$ for WIMP masses in the range $7-30$~GeV.

\section{Analysis strategy}

We use data collected during 14~months in $2009-2010$ by an array of ten ID detectors within the experimental setup described in detail in Ref.~\cite{edw2}. In each bolometer, individual interactions are detected by a thermal (heat) sensor, and the charge created during an interaction is collected by a set of six electrodes -- two fiducial electrodes, two veto and two guard channels. The electric field geometry created by the voltages applied to the electrodes isolates a central fiducial volume within the detector. Interactions within the fiducial volume produce equal amplitude and opposite polarity signals on the fiducial electrodes, while surface interactions induce signals also on the veto and guard electrodes~\cite{id}.

To be sensitive to recoil energies well below 20~keV and to minimize backgrounds in the WIMP search region, we rejected, a priori, six detectors among the ten bolometers available.
Four of these six detectors, which had nonfunctioning electrodes and/or poor resolutions for one or several channels, were rejected to guarantee the discrimination of the gamma-ray and surface event backgrounds around 10~keV.
Another detector was rejected due to the presence of a relatively intense accidental source of $^{210}$Pb in its vicinity, leading to a large number of low-energy and very low ionization-yield events induced by $\alpha$ radiation and Pb recoils. Finally, one detector was rejected because a low-energy gamma-ray background within its fiducial volume, which was 4 times larger than in other bolometers was observed~\footnote{However, it has been checked that including data from this detector does not significantly affect the resulting combined WIMP limit.}.

Periods of good data quality were selected in order to provide a homogeneous dataset for each of the four detectors used in the analysis. This time period selection was based on the measured baseline resolutions of each channel on an hour-by-hour basis. A large part of the corresponding exposure loss is due to noisy periods in the heat channels. That noise can be associated with certain readout channels that exhibit stronger than ideal coupling to the microphonic noise caused by the cryogenic equipment that was in operation during data taking.
Depending on the detector, the number of live days selected by these cuts ranges from 120 to 247 days. Based on the achieved thresholds on heat and ionization signals described below, it is expected that a significant fraction of the sensitivity to low-mass WIMPs will be driven by one of the detectors, called ID3, for which 197 live days were accumulated with average FWHM resolutions on the heat and fiducial ionization of 0.8~keV and 0.7~keV (electron recoil energy scale), respectively. The measured fiducial mass of these 400~g ID detectors is 160~g~\cite{idfirst}. This measurement was done using single-scatter events from cosmogenic activation gamma-ray lines with energies of 9.0 and 10.4~keV. The fiducial volume is primarily related to the electric field configuration inside the crystals, and the fiducial cuts are based on signals from veto electrodes, which do not depend on the fiducial energy. As a consequence, the measurement is applicable in the energy range considered for this study, and yields a total WIMP search exposure of 113~\kgdd.

The event reconstruction is identical to that carried out in~\cite{edw2}, as are standard event-based quality cuts. In particular, a cut is applied on the $\chi^2$ of the heat pulse reconstruction. The efficiency of this cut is independent of heat energy, and is 98.7\%, determined from a measurement using the aforementioned internal radioactivity lines at 10~keV. For the WIMP search we also reject events in coincidence with interactions recorded by any other surrounding bolometer in operation, or in the muon veto. The deadtime associated with this rejection is 0.2~\%. Finally, we restrict the WIMP search to measured recoil energies below 20~keV, making this data set independent from the one analyzed in~\cite{edw2}.

In order to search for nuclear recoil (NR) energy signals near the threshold, it is advantageous to evaluate the recoil energy of each event using an estimator with a very good energy resolution. We evaluate the recoil energy $E_r$ \textit{under the assumption that the event is due to a nuclear recoil} as in~\cite{cdms-lowmass} by inverting the following formula:

$$ E_{\rm h}  = \frac{E_r}{1+|V|/3} \left[ 1+ \frac{|V|}{3} Q_n(E_r)\right].$$

\noindent This describes the relationship between the heat energy $E_{\rm h}$, calibrated by gamma-ray interactions, and the recoil energy $E_r$, taking into account the heating of the detector associated with the charge drift under the polarization voltage $V$ in volts (Luke-Neganov effect~\cite{neganov,luke}). Here $Q_n(E_r) = 0.16\, {(E_r/{\rm keV})}^{0.18}$ is the nuclear recoil ionization quenching factor used in~\cite{martineau}, and validated down to 5~keV using neutron calibrations. This estimator provides a better energy resolution on $E_r$ for nuclear recoils than the estimator used in~\cite{edw2,cdms}. Note that for a fiducial electron recoil, $E_r$ overestimates the real energy by a factor $\sim 2$ for the typically applied voltages.

At low energy an efficiency loss appears due to the online trigger. Events are recorded based on upward fluctuations of the bolometer temperature measurement (filtered to remove noise). The trigger threshold was varied during data acquisition, depending on the noise conditions of the channel, and was recorded to disk. At any given time, the dependence of the trigger efficiency on $E_r$ can be described as a function of recoil energy as $\epsilon_{\rm online}(E_r) = 0.5 [1+ {\rm Erf}( (E_r-E_{\rm thresh}) / \sigma \sqrt{2})]$, where $E_{\rm thresh}$ and $\sigma$ are the recorded threshold and the measured resolution, in NR energy scale. This parametrization was validated using the flat, low-energy Compton plateau in gamma-ray calibration data. For the selected data, the average trigger efficiency is 78\% at 5.0~keV (NR scale) and 90\% at 6.3~keV.

We now consider signals from the ionization channels, which are exclusively used to reject the main backgrounds of nonfiducial interactions such as surface beta radioactivity, gamma-ray-induced interactions in the fiducial volume, and ionizationless events.

\begin{itemize}
\item Fiducial events are selected by requiring the absence of any signal on the veto and guard electrodes. Since a significant fraction of the ionization signal is always present on these channels for surface events, this cut efficiently rejects nonfiducial interactions down to low energy. Although the rejection performance depends on the amount of energy detected in the nonfiducial electrodes, the efficiency loss induced by the cut does not depend on energy. The effect of this loss is included in the measured fiducial mass~\cite{idfirst}.

\item The reconstructed nuclear recoil energy $E_r$, and the weighted fiducial ionization energy $E_i$, in electron recoil scale, are calculated for events that pass the selection described above.$\,E_i$ is the combination of fiducial electrode signals which results in the best ionization resolution. 
In the $(E_r,E_i)$ plane, the fiducial gamma-ray interactions are located in a band centered along the line $E_i = E_r(1+Q_n\,|V|/3)/(1+|V|/3)$, the width of which is determined by the heat and ionization resolutions. This background is rejected by selecting events below the band in which 95\% of the gamma-rays are located.

\item In WIMP search conditions, a large number of events were recorded with signals only on the heat channel. These ionizationless events have several sources: Pb recoils from surface alpha radioactivity, internal radioactivity of the NTD sensors, and non-Gaussian noise of electromagnetic and thermal origins. In order to reject efficiently these events, we take advantage of the redundancy of both fiducial electrode measurements and require the pulse fit algorithm to reconstruct signals from these two electrodes within 30~$\mu \rm s$ from each other in the recorded traces. In addition, only events with a weighted fiducial ionization energy $E_i$ larger than twice the corresponding FWHM are selected, the average value of which ranged from 1.4 to 1.9~keV, depending on the detector.
\end{itemize}

\begin{figure}
\begin{center} \includegraphics[width=0.5\textwidth]{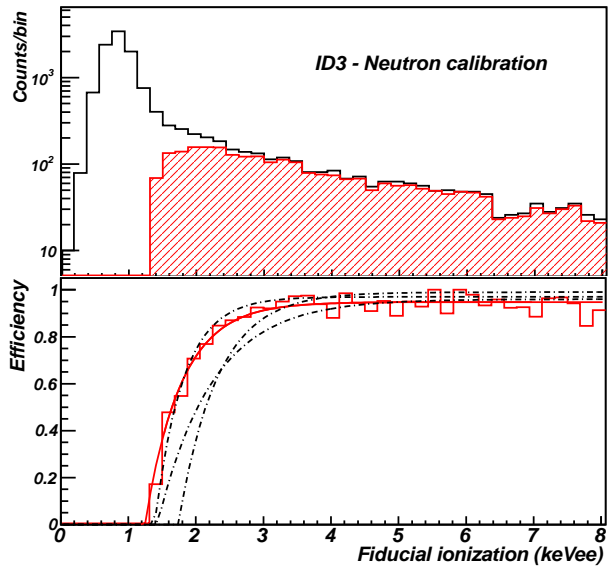} \end{center}
\caption{Top panel: Distribution of the fiducial ionization energy, $E_i$, in neutron calibration for the ID3 detector, before (black) and after (red) the ionization cut. The spectrum contains mostly nuclear recoils and a tail of ionizationless events. Bottom: cut efficiency determination. The fit function (continuous red line) is $\epsilon_{\rm ion} = 0.95 \left[ 1 - \exp(-1.87(E_{\rm ion} - 1.25))\right]$ where $E_{\rm ion}$ is in keVee (electron recoil scale). The dashed lines represent the efficiency functions estimated in the same way for the three other detectors used in this search.
\label{fig:ioneff}}
\end{figure}

\noindent The dependence of this last cut on the reconstructed fiducial ionization is measured from neutron calibrations, which provide a large sample of low-energy, WIMP-like pulses with ionization noise conditions identical to WIMP search conditions. Figure~\ref{fig:ioneff} presents the distribution of $E_i$ in neutron calibration for the ID3 detector, before and after this ionization cut. The measured efficiency function $\epsilon_{\rm ion}(E_i)$ is obtained through an analytic fit to the ratio of both histograms. Because of the presence of a residual tail of ionizationless events even in the short run time with a neutron source, the measured efficiency is underestimated at low energy, and therefore provides a conservative estimate of the WIMP sensitivity. 

The WIMP signal density is calculated based on the WIMP-induced nuclear recoil spectrum as a function of true recoil energy $p_0(E_{r0})$ as parametrized in~\cite{savage}.
For a given detector and a given WIMP mass, taking into account the measured resolutions of fiducial ionization energy $\sigma_i$ (electron recoil scale) and heat energy (NR scale) $\sigma_r$, the WIMP signal density in the $(E_r,E_i)$ plane is then:
\begin{eqnarray*}
\rho(E_r,E_i) & = & \frac{\epsilon_{\rm online}(E_r)\,\epsilon_{\rm ion}(E_i)}{2\pi\sigma_r \sigma_i} \int dE_{r0}\, p_0(E_{r0}) \\
&  \times &  {\rm exp} \left[ -\frac{(E_r-E_{r0})^2}{2\sigma_r^2} - \frac{(E_i-Q_n\,E_{r0})^2}{2\sigma_i^2} \right]. \\
\end{eqnarray*}
\noindent This function is normalized for a cross section $\sigma_{\rm SI}=10^{-6}$~pb and an exposure of 1~\kgdd, as it scales trivially with these parameters.
For each detector and each WIMP mass, we then define a WIMP search region in the $(E_r,E_i)$ plane as the region containing 90\% of all the calculated WIMP signal density below the gamma rejection cut, mentioned above. For example, Fig.~\ref{fig:qplot_id3} shows the function $\rho$ and the WIMP search region for ID3 and $M_{\chi}=10$~GeV.

The parameters $Q_n(E_r)$,  $\sigma_r$ and $\sigma_i$, entering in the definition of this WIMP search region, were cross-checked using calibrations down to the lowest relevant energies: the width of the gamma-ray band as well as the position and width of the neutron band are compatible with the measured baselines and with the parametrization $Q_n=0.16\,{(E_r/{\rm keV})}^{0.18}$ from~\cite{martineau}.

\section{Backgrounds, results and discussion}

Several backgrounds were anticipated :
\begin{itemize}
\item Ionizationless pulses were the most prominent background at very low energy, since we observe between 1000 and 5000 such events above 5~keV per detector. The rejection factor of the ionization cut is of the order of $\sim 10^{-6}$, resulting in a negligible contribution to the remaining background.
\item The residual fiducial gamma-ray background is estimated by extrapolation of the observed rate of gamma-ray events in the energy range for which the WIMP search region is limited by the 95\% gamma rejection cut. This range depends on the detector and WIMP mass under consideration. For example, for $M_{\chi}=10$~GeV its average value is $5.9 - 8.8$~keV. The estimated background from Gaussian leakage is then 2.5\% times the measured number of gamma events in that energy range, which leads to an overall estimate of $(1.2\pm0.2)$~events. For $M_{\chi}=30$~GeV this figure becomes $(1.1\pm0.2)$~events.
\item An upper limit on neutron backgrounds from different sources, as determined by the same combination of simulations and measurements as in~\cite{edw2}, was found to be 1.7 events in the energy range $5 - 20$~keV for the exposure of 113~\kgdd. A most probable contribution of 1.0 events is expected from radioactivity in the warm electronics, cables and connectors.
\item Surface interactions due to beta radioactivity are rejected but we have not measured rejection factors in the relevant $5 - 15$~keV energy range.
\end{itemize}

\begin{figure}
\begin{center} \includegraphics[width=0.5\textwidth]{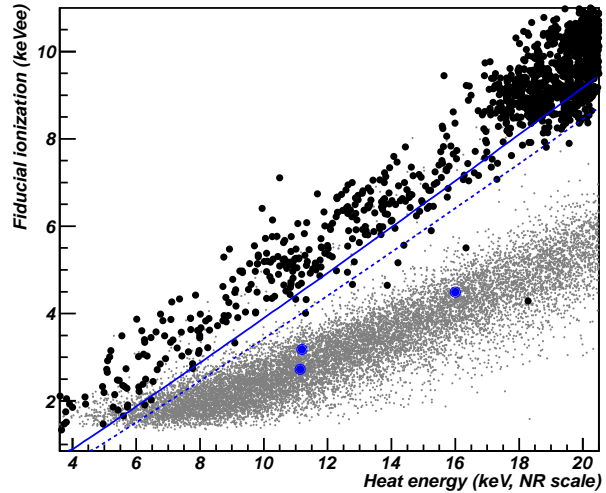} \end{center}
\caption{Scatter plot of heat vs ionization for all selected events in the 113~\kgd WIMP search. Neutron calibration data are also displayed as grey dots. The continuous (dashed) blue line represents the 95\% gamma-ray rejection cut for detector ID3 (ID401). The blue, circled points are events contained in the WIMP search region for at least some WIMP masses.
\label{fig:qplot}}
\end{figure}

\begin{figure}
\begin{center} \includegraphics[width=0.5\textwidth]{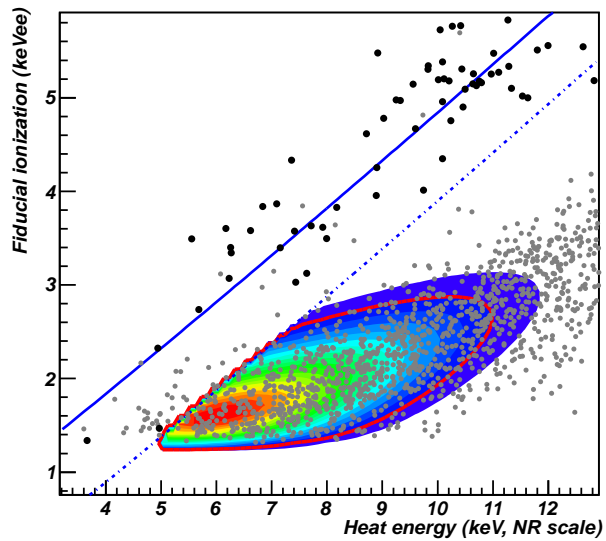} \end{center}
\caption{Color map of the WIMP signal density $\rho(E_r,E_i)$ together with the 90\% WIMP search box (red line) for a 10~GeV WIMP in detector ID3. Overlaid are background (black dots) and neutron calibration data (grey dots) for the same detector. The average position of the fiducial gamma-ray band and the corresponding 95\% rejection cut are shown as continuous and dashed blue lines, respectively.
\label{fig:qplot_id3}}
\end{figure}

Figure~\ref{fig:qplot} shows the $(E_r,E_i)$ scatterplot of the events recorded in the 113~\kgd exposure selected for the WIMP search. Most events are compatible with gamma-rays interacting in the fiducial volume. The distribution of $E_i$ for these events shows the presence of the cosmogenic activation lines at 9.0 and 10.4~keV, as well as less intense lines in the $5 - 7$~keV energy range. While a few events are present in the nuclear recoil region, there is no indication for an exponential distribution of nuclear recoils, in particular at energies below 10~keV, where most of the WIMP signal for $M_{\chi}\sim 10$~GeV is expected. Figure~\ref{fig:qplot_id3} shows the signal density $\rho(E_r,E_i)$ and the WIMP search region corresponding to $M_{\chi}=10$~GeV for the detector ID3, showing the absence of a nuclear recoil signal in spite of the sensitivity to such a signal demonstrated by neutron calibration data.

\begin{figure}
\begin{center} \includegraphics[width=0.5\textwidth]{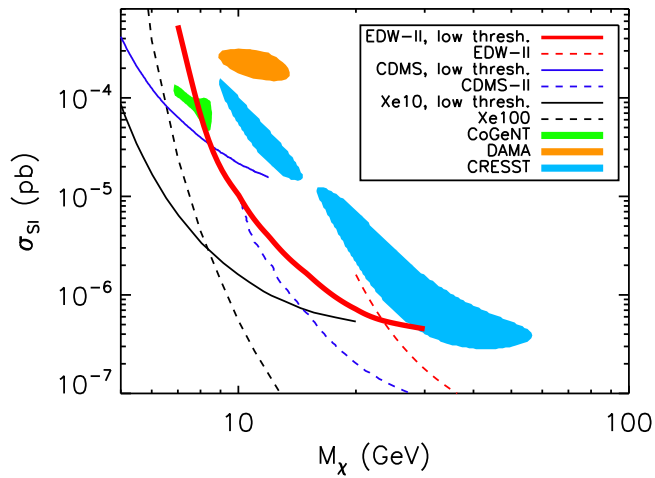} \end{center}
\caption{90\% CL Poisson limit on $\sigma_{\rm SI}$ as a function of WIMP mass derived from the analysis of the four bolometers (bold red line). We also show the location of potential WIMP signals from the CoGeNT~\cite{cogent}, CRESST~\cite{cresst} and DAMA~\cite{sav_dama} experiments, as well as constraints from EDELWEISS-II~\cite{edw2}, CDMS-II~\cite{cdms} and XENON100~\cite{xe100}, and the dedicated low-mass searches by CDMS-II~\cite{cdms-lowmass} and XENON10~\cite{xe-lowmass}.
\label{fig:limit}}
\end{figure}

The total residual background in WIMP search regions ranges from one event for $M_{\chi}=10$~GeV to three events at 30~GeV. This is compatible with the expected backgrounds described above. For relatively large masses, the sensitivity of this study is therefore background limited, while for low WIMP mass we are primarily limited by the achievable threshold.

To derive the corresponding limit on $\sigma_{\rm SI}$ as a function of WIMP mass, we count the events located within the WIMP search region defined previously, simply adding the number of events over the four selected detectors. The expected number of events for a given value of $\sigma_{\rm SI}$ is obtained by integrating the WIMP signal density $\rho(E_r,E_i)$ over the WIMP search region, and summing over the detectors. We then compute a 90\% CL limit on the total number of events using Poisson statistics. The obtained limit is shown as a function of $M_{\chi}$ for $7 < M_{\chi} < 30$~GeV in Fig.~\ref{fig:limit}.

This result extends the sensitivity of the previous EDELWEISS-II analysis down to WIMP masses below 20~GeV. We  exclude the entire zones corresponding to the interpretation of DAMA~\cite{dama} and CRESST~\cite{cresst} results in terms of elastic, spin-independent WIMP scattering. These zones extend down to a WIMP mass of 9~GeV, which is the "WIMP-safe" mass~\cite{pdg} where our experimental sensitivity is around 1~\% of the total WIMP signal recoil spectrum. While we significantly constrain part of the CoGeNT parameter space~\cite{cogent}, we cannot exclude the region corresponding to $M_{\chi} < 8$~GeV due to lack of sensitivity to nuclear recoil energies below 5~keV. 

Our results provide a cross-check of those obtained with other technologies, using dual-phased xenon TPCs (\cite{xe-lowmass,xe100}) or germanium detectors (\cite{cdms-lowmass,cogent}). In contrast to the latter, the WIMP search with EDELWEISS ID detectors presented in here is not limited by the presence of a large background as seen in CDMS and CoGeNT, which may be due to surface interactions. Although we do not have a precise measurement of the rejection factor for this background below 15~keV, the WIMP search data presented here demonstrates that it remains excellent down to 5~keV. This provides strong motivation for measuring this rejection factor with new-generation detectors. We aim to make further improvements in lowering energy thresholds and increasing the mass while maintaining a background-free region of interest for low-mass WIMP searches, including annual modulation measurements.

\section*{Acknowledgements}
The help of the technical staff of the Laboratoire Souterrain de Modane and the participant laboratories is gratefully acknowledged. This project is supported in part by the Agence Nationale pour la Recherche under contract ANR-10-BLAN-0422-03, by the Russian Foundation for Basic Research and by the Science and Technology Facilities Council, United Kingdom.

\end{document}